\documentclass[12pt]{article}
\usepackage{amsmath,amssymb}
\usepackage{color}
\usepackage{hyperref}

\makeatletter

\newcommand{\ie}{i.e.,~}

\@addtoreset{equation}{section}
\makeatother

\newcommand{\suppressed}{}
\newcommand{\modified}[1]{#1}

\begin{document}

\thispagestyle{empty}

\setcounter{page}{0}

\mbox{}
\vspace{5mm}

\begin{center} {\bf \Large  Remarks on Gauge Invariance\\[3mm] and First-class Constraints}

\vspace{1.5cm}

Marc Henneaux$^{a,b}$, Axel Kleinschmidt$^{a}$ and  Gustavo Lucena G\'omez$^{a}$

\footnotesize
\vspace{1 cm}

${}^a${\em Universit\'e Libre de Bruxelles and International Solvay Institutes, ULB-Campus Plaine CP231, 1050 Brussels, Belgium}

\vspace{.2cm}

${}^b${\em Centro de Estudios Cient\'{\i}ficos (CECS), Casilla 1469, Valdivia, Chile}\\

\vspace{7mm}
{\tt\{henneaux,axel.kleinschmidt,glucenag\}@ulb.ac.be}
\end{center}

\vspace {12mm}
\centerline{\bf Abstract}
\vspace{.2cm}
\noindent
Gauge symmetries lead to first-class constraints.  This assertion is of course true only for non trivial gauge symmetries, i.e., gauge symmetries that act non trivially on-shell on the dynamical variables.  We illustrate this \suppressed well\modified{-}appreciated fact for time reparametrization invariance in the context of modifications of gravity \modified{-- suggested in a recent proposal by Ho\v{r}ava --} in which the Hamiltonian constraint is deformed by arbitrary spatial diffeomorphism invariant terms, where some subtleties are found to arise.\footnote{To appear in the proceedings of the Conference ``Gauge Fields: Yesterday, Today, Tomorrow", dedicated to the 70th anniversary of Professor A.A.Slavnov.}

\newpage
\setcounter{footnote}{0}

\section{Introduction}

It is a pleasure to dedicate this paper to Andrei Alexeevich Slavnov on the occasion of his 70th birthday, wishing him continued success in his scientific career.

Gauge symmetries have been a central theme in Andrei Alexeevich's investigations.  This short note emphasizes a well-known fact (well\modified{-}known, but not always immediately apparent in some contexts) on the relation between gauge invariance and first-class constraints.  

The Hamiltonian formulation of theories with a gauge invariance has been given by Dirac \cite{Dirac:1950pj,Henneaux:1992ig}.  Gauge symmetries (i.e., symmetries i\modified{n}volving arbitrary functions of time) imply first-class constraints.  This has become one of the main lessons of Dirac's analysis.  However, implicit in this assertion is that the result holds only for non trivial gauge symmetries \modified{--} as it should.

\section{Trivial Gauge Symmetries}

Trivial gauge symmetries are symmetries that do not act on the dynamical variables when the equations of motion hold. 

An elementary example is provided by the following transformation
\begin{equation}
\delta q = 2 \epsilon \dddot{q} + \dot{\epsilon} \ddot{q}\modified{,} \label{fake1}
\end{equation}
where $\epsilon(t)$ is an arbitrary function of time.  This transformation leaves the free non-relativistic particle action
\begin{equation}
S[q(t)] = \frac{1}{2} \int  \modified{\text{d}}t \, \dot{q} ^2
\end{equation} invariant (up to a surface \modified{term}) for any $\epsilon(t)$.

However, the existence of this invariance has no dynamical implication: it leads to no first-class constraint and no ambiguity in the general solution to the equations of motion for given initial data $(q_0, \dot{q}_0)$.  Indeed, the gauge invariance (\ref{fake1}) vanishes ``on-shell'', i.e., when the equations of motion $\ddot{q}=0$ hold.  When rewritten in terms of phase space variables, its canonical generator identically vanishes.  For this reason, the gauge symmetry (\ref{fake1}) can safely be ignored and one can live in peace without even mentioning it.  It is called an ``\modified{on}-shell trivial gauge symmetry" or, for short, a ``trivial gauge symmetry".

Quite generally, for a system with several degrees of freedom $q^i$ and action $S[q^i]$, on-shell trivial gauge symmetries take the form
\begin{equation}
\delta q^i =  \mu^{ij} \frac{\delta S}{\delta q^j}\modified{,} \label{fake2}
\end{equation} 
where $\mu^{ij} = - \mu^{ji}$ is antisymmetric but otherwise arbitrary and where we have used \modified{DeWitt}'s condensed notations in which the sum over $j$ implies also an integration over time (and space in the case of field theory).  Explicity in the case of classical mechanics,
\begin{equation}
\delta q^i (t) =  \int \modified{\text{d}}t' \, \mu^{ij}(t,t')\,  \frac{\delta S}{\delta q^j(t')} \modified{,}
\end{equation}
where  $\mu^{ij}(t,t') = - \mu^{ji}(t',t)$.

In the case of (\ref{fake1}), one has
\begin{equation}
\delta q(t) =  \int \modified{\text{d}}t' \, \mu(t,t')\,  \frac{\delta S}{\delta q(t')} \modified{,}
\end{equation}
with 
\begin{equation}
\mu(t,t') = - \left( \epsilon(t) + \epsilon(t')\right) \frac{\modified{\text{d}}}{\modified{\text{d}}t} \delta(t-t')\modified{.}
\end{equation}
It is in fact a general theorem that under reasonable regularity assumptions on the action, any symmetry of $S$ that vanishes on-shell takes the form (\ref{fake2}) \cite{Henneaux:1992ig}(section~3.1.5).

Trivial gauge symmetries can arise in the commutator of non-trivial symmetries.  This occurs in supergravity \modified{and} it is in this way that their existence was uncovered in modern field theory.

\section{Another example}

Another example, more adapted to our purposes, is given by the model with action
\begin{equation} S[q^i, p_i, N^\alpha] = \int \text{d}t \left( p_i \dot{q}^i - N^\alpha {\mathcal H}_\alpha \right).  \label{ClassMechModel}
\end{equation}
The action (\ref{ClassMechModel}) is clearly invariant under time reparametrizations
\begin{align}
 \delta q^i &= \eta \dot{q}^i, & \delta p_i &= \eta \dot{p}_i, & \label{delta1}\\
 \delta N^\alpha &= (\eta N^\alpha)\dot{ }\,, & \delta {\mathcal H}_\alpha &= \eta\dot{\mathcal H}_\alpha, \label{delta2}
\end{align}
independently of the nature of the constraints $ {\mathcal H}_\alpha \approx 0$, where $\eta(t)$ is an arbitrary function of time (the variation of ${\mathcal H}_\alpha$ being a consequence of (\ref{delta1})).  This simply follows from tensor calculus in one dimension, the $q$'s and the $p$'s being scalars and the multipliers $N^\alpha$ being densities of weight one.

Now, it would seem that a gauge symmetry as  ``respectable" as time reparametrization \suppressed cannot be trivial.  However, this depends on the nature of the constraints.  If all the constraints are second-class, the equations of motion imply $N^\alpha = 0$ from $\dot{\mathcal H}_\alpha = 0$ and hence also $\dot{q}^i = 0$, $\dot{p}_i = 0$.  Accordingly, the variations of all the variables $q^i$, $p_i$ and $N^\alpha$ in (\ref{delta1}), (\ref{delta2}) vanish on-shell.  This implies that  diffeomorphism invariance is \modified{in this case} a trivial symmetry.

To exhibit this fact explicitly, we observe that when the constraints are all second-class, the matrix $C_{\alpha \beta}$ defined by  $[{\mathcal H}_\alpha, {\mathcal H}_\beta] \equiv C_{\alpha \beta}$ is invertible,  $\det(C_{\alpha \beta}) \not= 0$.  We denote the inverse matrix by $C^{\alpha \beta}$, so that $C^{\alpha \rho} C_{\rho \beta} = \delta^\alpha_\beta $.  Then the diffeomorphism transformations can identically be rewritten as antisymmetric combinations of the equations of motion \suppressed
\begin{eqnarray}
&& \delta q^i = \eta \left[ \frac{\delta S}{\delta p_i} + \frac{\partial {\mathcal H}_\rho}{\partial p_i} C^{\rho \alpha} \left(- \frac{\text{d}}{\text{d}t} \frac{\delta S}{\delta N^\alpha} - \frac{\partial {\mathcal H}_\alpha}{\partial q^j} \frac{\delta S}{\delta p_j} + \frac{\partial {\mathcal H}_\alpha}{\partial p_j} \frac{\delta S}{\delta q^j} \right) \right]\modified{,} \\
&& \delta p_i = \eta \left[ - \frac{\delta S}{\delta q^i} - \frac{\partial {\mathcal H}_\rho}{\partial q^i} C^{\rho \alpha} \left(- \frac{\text{d}}{\text{d}t} \frac{\delta S}{\delta N^\alpha} - \frac{\partial {\mathcal H}_\alpha}{\partial q^j} \frac{\delta S}{\delta p_j} + \frac{\partial {\mathcal H}_\alpha}{\partial p_j} \frac{\delta S}{\delta q^j} \right) \right]\modified{,} \\
&& \delta N^\alpha = \frac{\text{d}}{\text{d}t}\left[ \eta C^{\rho \alpha} \left(- \frac{\text{d}}{\text{d}t} \frac{\delta S}{\delta N^\alpha} - \frac{\partial {\mathcal H}_\alpha}{\partial q^j} \frac{\delta S}{\delta p_j} + \frac{\partial {\mathcal H}_\alpha}{\partial p_j} \frac{\delta S}{\delta q^j} \right) \right].
\end{eqnarray}
Note that
\begin{equation}
N^\rho = C^{\rho \alpha} \left(- \frac{\text{d}}{\text{d}t} \frac{\delta S}{\delta N^\alpha} - \frac{\partial {\mathcal H}_\alpha}{\partial q^j} \frac{\delta S}{\delta p_j} + \frac{\partial {\mathcal H}_\alpha}{\partial p_j} \frac{\delta S}{\delta q^j} \right) .
\end{equation}
We stress that the appearance of the inverse $C^{\alpha\beta}$ clearly signals that this argument only holds if all the constraints are second-class. 

Thus, we conclude that one cannot decide beforehand whether the gauge symmetry (\ref{delta1}), (\ref{delta2}) is trivial or non-trivial, whatever one's geometrical prejudices are.  To determine the nature of the gauge symmetry requires a more detailed dynamical analysis.

\section{Deformations of gravity theory}

\subsection{The models}

Recently, Ho\v{r}ava proposed a candidate for a UV completion of Einstein theory of gravity in which full spacetime diffeomorphism invariance is abandoned and recovered only at large distances \cite{Horava:2008ih,Horava:2009uw}.  Based on appealing analogies with condensed matter physics and anisotropic scaling  \modified{\emph{\`a la}} Lifschitz (see~\cite{Horava:2008ih,Horava:2009uw} and references therein), it has been proposed that this alternative to Einstein theory might provide a renormalizable UV completion of general relativity and therefore  yields a very attractive approach that is worth being explored.

This suggests studying deformations of general relativity in which one keeps intact the momentum constraints, but deforms the Hamiltonian constraint by arbitrary terms compatible with spatial diffeomorphism invariance.

We thus consider theories described by the following Hamiltonian data:
\begin{itemize}
\item Phase space variables, $g_{ij}(x)$, $\pi^{ij}(x)$, with Poisson brackets
\begin{equation}
{}[g_{ij}(x), \pi^{mn}(y)]= \frac{1}{2}(\delta^{m}_{i}\delta^{n}_{j}+\delta^{n}_{i}\delta^{m}_{j})\delta(x,y),
\end{equation}
where $x$ and $y$ are points on a spatial slice and $\pi^{ij}$ is the momentum conjugated to the spatial metric $g_{ij}$.
\item Constraint surface in phase space defined by constraints
\begin{align}
\label{hamconstraints}
 {\mathcal H}(x) &\approx 0   \quad\quad&& \hbox{ (``Hamiltonian constraint"),}\\
 {\mathcal H}_k(x) &\approx 0  \quad\quad&&\hbox{ (``momentum constraints")}\modified{,}
\end{align}
where weak equalities $\approx$ mean zero on the constraint surface, as usual.
\item Equations of motion generated by a Hamiltonian
\begin{equation}\label{ham}
H = \int \text{d}^3x \left( N(x) \, {\mathcal H}(x) + N^k(x) \, {\mathcal H}_k(x) \right)\modified{,}
\end{equation}
with lapse function $N(x,t)$ and shift vector $N^k(x,t)$.\footnote{Very often, the time dependence is not written explicitly.} \modified{From the point of view of Ho\v{r}ava gravity, we are thus in the non-projectable class of theories.\footnote{Our insistence on this form of the Hamiltonian, with $N$ and $N^k$ appearing as Lagrange multipliers, is modeled on the canonical structure of general relativity. Our analysis therefore does not cover further modifications of Ho\v{r}ava gravity, discussed for example in~\cite{Blas:2009qj,Kluson:2010xx,Kluson:2010nf}.}}
\end{itemize}
The action is
\begin{equation}
S[g_{ij},\pi^{ij},N,N^k] = \int \text{d}t \left[\left(\int \text{d}^3x \, \pi^{ij} \dot{g}_{ij}\right) - H \right]. \label{action0}
\end{equation}
The equations of motion follow by extremizing the action with respect to $g_{ij}(x)$ and  $\pi^{ij}(x)$ (dynamical equations of motion), as well as with respect to the lapse and the shift functions that serve as Lagrange multipliers for the constraints. They are given by
\begin{equation}
\dot{F} = [F,H]
\end{equation}
for any function(al) $F[g_{ij}(x),\pi^{ij}(x)]$ of the canonical variables, together with the above constraints.  

 \modified{{}From a technical point of view, one may regard the constraints as the secondary constraints resulting from the preservation in time of the primary constraints expressing that the conjugate momenta  $\pi_N$ and $\pi_{N^k}$ to the lapse and the shift are zero. We have used these primary constraints to eliminate $\pi_N$ and $\pi_{N^k}$.  Whether there are further constraints must be analysed through the application of the Dirac algorithm. For instance, if the brackets between the Hamiltonian and momentum constraints are all zero on the constraint surface, these constraints are first-class. There are no further constraints and the multipliers $N$, $N^k$ are undetermined.  This is what happens in general relativity. At the other extreme, if the bracket matrix $[{\mathcal H}(x), {\mathcal H}(y)]$ is invertible, the Hamiltonian constraints are second-class and $N$ is completely determined (while ${\mathcal H}_k$ remains first class and $N^k$ remains undetermined, see below).  In that case, there are also no additional constraints.  Whether one encounters the first or the second case depends on the values of the coupling constants.  It is the objective of the remaining of this paper to analyse this point further.  As we shall see and comment in the sequel, the situation is in fact more complicated because the rank of $[{\mathcal H}(x), {\mathcal H}(y)]$ is not constant on the constraint surface.}

\subsection{Form of constraints}

The form of the momentum constraints, which generate spatial diffeomorphisms, is universal and given by
\begin{equation}
{\mathcal H}_k = - 2 \nabla_i \pi^i_{\; \; k}\modified{,}
\end{equation}
where $\nabla$ stands for the spatial covariant derivative operator. Indices are lowered and raised with the spatial metric $g_{ij}$ and its inverse $g^{ij}$.

By contrast, the Hamiltonian constraint depends on various coupling constants.  We only require that ${\mathcal H}$ be a density of weight one under spatial diffeomorphisms and be quadratic in the momenta, so that it takes the form
\begin{equation}
{\mathcal H} = {\mathcal H}_1 + {\mathcal H}_2\modified{,}
\end{equation}
where ${\mathcal H}_1$ is the kinetic term\footnote{\modified{Here,} $\lambda$ is the parameter appearing in the modified DeWitt metric on the space of metrics and is expected to go to zero in the IR limit if general relativity is to be recovered at low energies (see \cite{Horava:2009uw} for the details).} (with $\pi = g_{ij}\pi^{ij}$)
\begin{equation}
{\mathcal H}_1 = \frac{1}{\sqrt{g}}\left(\pi^{ij} \pi_{ij} - \frac{\lambda}{3 \lambda - 1} \pi^2 \right)
\end{equation}
and ${\mathcal H}_2$ contains the potential terms
\begin{equation}\label{H2}
{\mathcal H}_2 = \sqrt{g} \left( \sigma + \xi R + \eta R^2 + \zeta R^{ij}R_{ij} +\beta C_{ij}C^{ij}+\gamma R \triangle R + \ldots\right).
\end{equation}
Here, the spatial Laplacian is $\triangle = \nabla^i \nabla_i$. 
One may impose the further restriction that ${\mathcal H}$ contains at most six  \modified{spatial} derivatives of the metric so that it is formally power-counting renormalizable \cite{Horava:2008ih,Horava:2009uw}. However, this restriction is not necessary for the general analysis of the consistency of the system which is given below, and any invariant constructed out of the spatial curvature and its successive covariant derivatives is allowed.  

General relativity corresponds to the choice $\lambda = 1$ and all other coupling constants equal to zero except $\sigma$ and $\xi$.   Switching on the other couplings yield deformations of general relativity.  A notable choice, different from general relativity, is the ``ultralocal" theory of \cite{Teitelboim:1981fb,Isham:1975ur,Henneaux:1979vn}, which has all couplings equal to zero but $\lambda$ and $\sigma$.  This theory possesses the same number of gauge invariances and degrees of freedom as general relativity, which, as we shall see, makes it rather special among the deformations.

For any values of the coupling constants, the constraints  ${\mathcal H}_k(x)$,  ${\mathcal H}(x)$ fulfill the algebra
\begin{eqnarray}
&&[{\mathcal H}_k(x),{\mathcal H}_m(x')] = {\mathcal H}_k(x') \delta_{,m}(x-x') + {\mathcal H}_m(x)  \delta_{,k}(x-x'),   \\
&&[{\mathcal H}(x),{\mathcal H}_k(x')]  =  {\mathcal H}(x')  \delta_{,k}(x-x')
\end{eqnarray}
expressing that \suppressed ${\mathcal H}_k(x)$ are the generators of spatial diffeomorphisms and that ${\mathcal H}(x)$ is a density of weight one.
It clearly follows from the constraint algebra that the constraints ${\mathcal H}_k(x)$ are first-class.  Whether there are further first-class constraints depend on the values of the coupling constants.

\subsection{Time reparametrization invariance}

In addition to being invariant under arbitrary spacetime-dependent spatial diffeomorphisms,
\begin{subequations}
\label{spacerep}
\begin{align}
\delta g_{ij} &= \eta^kg_{ij\; ,k} + \eta^k_{\;\; ,i}g_{kj} + \eta^k_{\; \; ,j}g_{ik} , \\
\delta \pi^{ij} &= (\eta^k \pi^{ij})_{ ,k} - \eta^i_{\; \; ,k} \pi^{kj} - \eta^j_{\; \; ,k} \pi^{ik}, \; \; \\
\delta N &= \eta^k N_{, k} , \\
\delta N^i &= \dot{\eta}^i + \eta^k N^i_{\;\; ,k} - \eta^i_{\; \; ,k}N^k ,
\end{align}
\end{subequations}
the action (\ref{action0}) is also invariant under space-independent time reparametrizations $\eta(t)$ for any choice of the coupling constants,
\begin{subequations}
\label{timerep}
\begin{align}
\delta g_{ij} &= \eta \dot{g}_{ij}, \label{TR1}\\
\delta \pi^{ij} &= \eta \dot{\pi}^{ij}, \label{TR2}\\
\delta N &= (\eta N)\dot{} \, , \label{TR3}\\
\delta N^k &= (\eta N^k)\dot{}\, .\label{TR4}
\end{align}
\end{subequations}

To the spatial diffeomorphisms correspond the first-class constraints \linebreak ${\mathcal H}_k\modified{(x)} \approx 0$, as we already pointed out.   One might be tempted to infer from time reparametrization invariance that at least one combination of the  constraints ${\mathcal H} \approx 0$ should be first-class, for any choice of the coupling constants.  That combination would be the canonical generator of the symmetry. However, as we pointed out above, there is no such guarantee since time reparametrizations might be trivial.  Settling this question requires a more detailed, direct analysis of the dynamics.  

\section{Time reparametrization  \suppressed is generically trivial for the above class of deformations}
\subsection{Matrices depending continuously on parameters}

Before tackling this question, we recall an important property of matrices depending on parameters.

Let $M(\beta^{\modified{A}})$ be a $N \times N$ matrix depending continously on parameters $\beta^A$.  Assume that $M$ is invertible for some values $\beta^A_0$ of the parameters upon which it depends.  Then, because $\det M \not= 0$ is an inequality, the matrix $M$ is also invertible in an open neighbourhood of $\beta^A_0$ and is in that sense ``generically invertible".  The condition $\det M = 0$, being a non trivial equation (non\modified{-}trivial because there are values of the parameters for which $\det M \not= 0$), defines by contrast a subset of lower dimension in the space of parameters.  Values of the parameters for which $\det M = 0$ are `` non\modified{-}generic".

This crucial property is strictly speaking valid only for finite-fimensional matrices.  We shall however proceed as if it were also true for the infinite-dimensional matrix $G(x,y)$ that appears in the analysis below.

\subsection{The matrix $G(x,y) = [ {\mathcal H}(x),  {\mathcal H}(y)] $ is generically invertible}

Requesting that the constraint surface be preserved by the dynamics, \ie $$\dot{{\mathcal H}}(x) = [{\mathcal H}(x), H] =  \int \text{d}^3 y \, G(x,y) N(y) \approx 0, $$ with 
\begin{equation}
G(x,y) = [ {\mathcal H}(x),  {\mathcal H}(y)] , 
\end{equation}
leads to a partial differential equation (in space) for the lapse function $N$ of the form \begin{equation}
\alpha^{ijkl} \nabla_{ijkl} N + \beta^{ijk} \nabla_{ijk} N + \gamma^{ij} \nabla_{ij} N + \delta^{i} \nabla_{i} N + \omega N\approx 0 \modified{,} \label{key}
\end{equation}
where $\alpha^{ijkl} = \alpha^{(ijkl)} $, $ \beta^{ijk} =  \beta^{(ijk)}$, $\gamma^{ij} = \gamma^{ji}$, $\delta^i$ and $ \omega$ are functions of the canonical variables that depend on the coupling constants and $\nabla_{ij} =\nabla_{(i}\nabla_{j)}$ etc. The explicit form of the coefficients will not be needed here. If terms containing higher derivatives of the metric (beyond order 6) are allowed in ${\mathcal H}$, then, there are higher derivatives of $N$ in (\ref{key}).  No matter what terms are included, this equation is linear homogeneous in $N$ and always admits the solution $N= 0$.

Analyzing the (co)rank of the matrix $G(x,y)$ is equivalent to determining on how many arbitrary constants does the general solution to equation (\ref{key}) depends.  If the only solution is $N=0$, the corank of $G(x,y)$ is zero and $G(x,y)$ is formally invertible.  Thus, the problem is to find the general solution of (\ref{key}) for arbitrary values of the coupling constants.

This is not an easy task because the coefficients in (\ref{key}) are complicated functions of the canonical variables and their derivatives.  That is, the matrix $G(x,y)$ depends  not only on the coupling constants but also on the values of the metric and its conjugate momentum (subject to the constraint equations),
$$ G(x,y)[\lambda, \sigma, \xi, \cdots, g_{ij}(z), \pi^{km}(z')].$$
This dependence is such that the rank itself of $G(x,y)$ also depends on the coupling constants and on the values of the metric and its conjugate momentum.  For instance, it is easy to verify that if $\pi^{ij}=0$ (which is compatible with the Hamiltonian constraint), then the equation (\ref{key}) degenerates to $0=0$ and imposes no condition on $N$. However, for other values of the canonical variables, (\ref{key}) is generically a non\modified{-}trivial equation.

In \cite{HKL}, we have analyzed (\ref{key}) for generic values of the coupling constants and of the canonical variables (subject to the constraint equations).  We have explicitly picked out values of the coupling constants, and, for this choice of the coupling constants, of the canonical variables fulfilling the constraint equations, so that the only solution of (\ref{key}) is $N= 0$.  This was done in the asymptotically flat context, imposing that $N$ should go to a constant at infinity.

This means that $G(x,y)$ is generically of maximal rank, i.e., invertible. Though at first sight contrary to intuition since the theory is time reparametrization invariant, there is no contradiction as time reparametrization becomes trivial when $N=0$, as we explained above.

We stress again the importance of the term ``generically" here. It means first of all ``generically" in parameter space.  For instance, the choices of parameters corresponding to general relativity or the ultralocal theory make $G(x,y)$ identically zero on the constraint surface and are not generic.  But also, for a given generic  choice of parameters, on must consider generic values of the canonical variables on the constraint surface.  As mentioned above, the static case with $\pi^{ij} =0$, much considered in the literature \cite{Lu:2009em,Kiritsis:2009rx,Kiritsis:2009vz}  is non\modified{-}generic and allows $N \not=0$.  Similarly, if the spatial sections have zero curvature, or are of constant curvature and the extrinsic curvature is a (time-dependent) multiple of the metric, as it is relevant for cosmological models \cite{Calcagni:2009ar,Kiritsis:2009sh,Bakas:2009ku}, the covariant derivatives of the spatial Riemann tensor and of the extrinsic curvature vanish so that $G(x,y)$ is also zero. These special cases of measure zero are blind to the restrictions on $N$ imposed by (\ref{key}).  

The choice of coupling constants that make the analysis tractable has been considered first in the context of Ho\v{r}ava's theory in \cite{Blas:2009yd} and earlier in a different context in \cite{Giulini:1994dx,barbour:2002}.
It is obtained by setting all coupling constants equal to zero, except $\lambda$ and $\xi$, with $\lambda \not=1$ in order to depart from general relativity.  The equation (\ref{key}) \modified{then reduces} to
\begin{equation}
(\lambda - 1) \nabla_i(u \nabla^i \pi)  \approx 0  \; \; \; \;  \Rightarrow \; \; \; \;  \nabla_i(u \nabla^i \pi)  \approx 0 \label{key4}
\end{equation}
with $u = N^2$.  This equation is \modified{non-trivial} for generic choices of $\pi$ (which is not restricted by the constraint equation) and implies $N=0$ in the class of functions that go to a constant at infinity \cite{HKL}.\footnote{The result $N=0$ is not mathematically inconsistent (i.e. not of the form $1 = 0$, see also below). Excluding it does not follow from Dirac algorithm but would be the consequence of postulates without clear geometrical origins in Riemannian geometry.}
 Of course, if $\pi=0$, the equation (\ref{key4}) implies no restriction on $N$, but this is  \modified{a} non\modified{-}generic  \modified{situation} on the constraint surface defined by ${\mathcal H} = 0$, ${\mathcal H}_k = 0$.  

\section{Conclusions}

One of the beauties of general relativity is that it is difficult to deform it without running into inconsistencies.  We have illustrated this property in the context of deformations of the Hamiltonian constraint by terms that are requested to preserve only spatial diffeomorphism invariance.  We have shown then that all the Hamiltonian constraints (for all $x$'s) are generically second-class, implying that the lapse is zero.  There is no contradiction with time reparametrization invariance, because this invariance then turns out to be an ``on-shell trivial" gauge symmetry with no physical implication.  

To illustrate this absence of contradiction due to the triviality of the gauge symmetry was the main motivation of this note.  We shall close our paper by making some further comments on the viability of the  deformations of general relativity considered here.

\subsection{Mathematical consistency versus physical considerations}

The fact that the Hamiltonian constraints are second-class is not, in itself, a mathematical inconsistency.  It simply tells us that the lapse is uniquely fixed, and, because the equation for $N$ is a homogeneous equation always possessing $N=0$ as a solution, it means that $N = 0$.  The theory then possesses $5/2$ degrees of freedom per space point since there are $6$ conjugate pairs, $3$ first-class constraints and one second-class constraint (per space point). We thus agree with reference \cite{Blas:2009yd}, which also concluded that the Hamiltonian constraints were (generically) second-class and determined the lapse. Earlier work on the difficulties of deformations of the Hamiltonian constraints are \cite{barbour:2002,XX,Charmousis:2009em,Li:2009bg}.

The extra $1/2$ degree of freedom (the so-called ``extra mode") might be thought of as contained in the pair formed by $\pi$ and the conformal factor (and not in $N$, which is identically zero).  The conformal factor is determined by the Hamiltonian constraints.  In general relativity, where the constraints are first-class, one uses the corresponding gauge freedom to impose a gauge condition on the conjugated $\pi(x)$. Here, the constraints are second-class, thus expressing instead that $\pi(x)$ is self-conjugate in the corresponding Dirac bracket (whose expression is rather intricate and will not be worked out here). We note that the extra mode is somewhat analogous to a chiral boson \cite{Floreanini:1987as,Henneaux:1988gg}, for which there is also a single second-class constraint per space point.

Since $N = 0$, the dynamics is very simple: the Hamiltonian vanishes (in the gauge where the shift is zero) and any function of the canonical variables is a constant of motion.  This is mathematically consistent but the theory not only differs in a drastic way from general relativity but is also physically rather meaningless as there is no time evolution.  One can therefore say that there is a dynamical inconsistency with what one requests from the theory on physical grounds, \ie the lapse should be non\modified{-}ero and belong to a one-parameter family of solutions (away from the general relativity values).

Although we have not investigated the equation for the lapse in the compact case, one might anticipate that difficulties in the analysis will also arise in that case since the solutions must be globally well-defined. Locality requirement for the lapse as a function of the other variables should presumably also be imposed in order to be able to apply the methods of local quantum field theory.  This appears to be also a very restrictive condition.  It is not unreasonable to expect similar difficulties with other asymptotic boundary conditions (e.g.,  anti-de Sitter spaces).

\subsection{Going beyond the above class of deformations}

That the only solution for the lapse is generically $N=0$ is clearly unsatisfactory from a physical point of view so that one must go beyond the class of deformations of general relativity considered here to get a physically consistent theory.   Some of these possibilities were already indicated in the original work \cite{Horava:2008ih,Horava:2009uw}.

\subsubsection{Non\modified{-}generic values of the coupling constants}

Our results do not exclude special values of the coupling constants for which extra non trivial gauge symmetry would be present. We know that all the Hamiltonian constraints are first-class for the choices corresponding to general relativity and the ultralocal theory.  Are there other (``non generic") choices of the coupling constants for which all the constraints are first-class? The recent analysis of \cite{Farkas:2010dw} gives a negative answer.  This does not exclude, however,  the possibility that there exist values of the coupling constants for which some (but not all) of the ${\mathcal H}(x)$'s are first-class.  To our knowedge this is an open question.

\subsubsection{Extra constraints}

Instead of searching for non\modified{-}generic values of the coupling constants that would enlarge the gauge symmetry, one might ``go non-generic on the constraint surface", i.e., further restrict phase space by imposing additional constraints.  This is not \modified{generically} a consequence of the application of Dirac's method since $N=0$ is mathematically consistent.  However, it is a possibility which is present here because the rank of the matrix $G(x,y)$ is not a constant over the constraint surface (a phenomenon investigated earlier in a different context in \cite{Banados:1995mq,Banados:1996yj,Banados:1997qs}). One might wish to exploit this possibility in order to avoid the disappointing result $N=0$.  \modified{The additional constraints would appear as ``tertiary constraints" emerging at non generic points on the constraint surface defined by the secondary constraints.}

For example, for the particular values of the couplings mentioned above ($\lambda\neq 1$, other couplings set to zero except $\xi$), one might impose the extra condition $\pi = 0$ as it was observed in \cite{HKL}.  The equation (\ref{key4}) implies then no condition on the lapse since it reduces to $0=0$.  This is a consistent subsector  \modified{which in this case} turns out in fact to be a gauge-fixed version of vacuum general relativity \cite{HKL} (if $\pi = 0$, one may redefine ${\mathcal H}$ by adding arbitrary multiples of $\pi^2$ and so set $\lambda = 1$). This case was later studied also in~\cite{Pons:2010ke,Bellorin:2010je}.

Whether this procedure is possible for different values of the coupling constants without reducing too much the theory remains to be seen (imposing $\pi^{ij} = 0$,  for instance, allows $N \not= 0$ but is too big a restriction, leaving too few degrees of freedom).  The analysis of \cite{Li:2009bg} seems to point out that this is impossible for generic values of the coupling constants.  If true, imposing further constraints would not be in general a satisfactory way to avoid $N=0$.  From that point of view, the obvious fact that imposing $\pi =0$ with the above particular choice of couplings reproduces vacuum general relativity would then not be representative of the general situation and would therefore be somewhat anecdotical.

\subsubsection{Different modifications}

Different strategies might be envisioned. For instance, one might consider different types of anistropic scalings in which the spatial dimensions are not all on the same footing \cite{Horava:2008ih,Horava:2009uw}. Or one might render the Lagrange multiplier $N$ dynamical (and non-zero) by allowing terms that are non linear in $N$ and its derivatives, dropping the above form of the Hamiltonian constraint \cite{Blas:2009qj}.  Whether these original possibilities lead to viable modifications of general relativity fulfilling the dream of being renormalizable remains to be seen.

\section*{Acknowledgements} Our work is partially supported by IISN - Belgium (conventions 4.4511.06 and 4.4514.08)
and by the Belgian Federal Science Policy Office through the Interuniversity Attraction Pole P6/11. AK is a Research Associate of the Fonds de la Recherche Scientifique--FNRS, Belgium.

\end{document}